\def\beq{\begin{equation}}
\def\eeq{\end{equation}}
\def\beeq{\begin{eqnarray}}
\def\beeqn{\begin{eqnarray*}}
\def\eeeq{\end{eqnarray}}
\def\eeeqn{\end{eqnarray*}}
\def\K{{\cal K}}
\def\Z{{\cal Z}}
\def\no{\nonumber}
\def\W{{\cal W}}

\documentstyle[preprint,aps,floats,amssymb]{revtex}
\begin{document}
\input{epsf}
\draft
\newfont{\form}{cmss10}
\title{Quantum 't Hooft loops of  SYM ${\cal N}=4$ as instantons
of YM$_2$ in dual groups SU(N) and SU(N)/Z$_{\rm N}$} 
\author{Antonio Bassetto, Shiyamala Thambyahpillai}
\address{Dipartimento di Fisica "G. Galilei" and
INFN, Sezione di Padova,\\
Via Marzolo 8, 35131 Padua, Italy,\\
e-mail: {\rm bassetto@pd.infn.it, thamby@particle.uni-karlsruhe.de}}
\maketitle
\begin{abstract}
A relation between circular 1/2 BPS 't Hooft operators in 
4d ${\cal N}=4$ SYM and instantonic solutions in 2d Yang-Mills theory
(YM$_2$) has recently  been conjectured. 
Localization indeed predicts that those
't Hooft operators in a theory with gauge group $G$ are captured by 
instanton contributions to the partition function of YM$_2$, belonging
to representations of the dual group $^LG$. This conjecture has been 
tested in the case $G=U(N)=$$^LG$ and for fundamental representations.
In this paper we examine this conjecture for the case of the 
groups $G=SU(N)$
and $^LG=SU(N)/Z_N$ and loops in different representations. Peculiarities
when groups are not self-dual and representations not ``minimal'' are
pointed out.
\end{abstract}
\vskip 1.0truecm
\noindent

\vskip.2truecm
DFPD/TH 15-2010.

\section{Introduction}
Electric-magnetic duality in electromagnetism \cite{dir} has been 
extended to non-Abelian theories  and,
in particular, to ${\cal N}=4$ super Yang-Mills (SYM ${\cal N}=4$), 
(S-duality)\cite{mon}.
It is conjectured that SYM ${\cal N}=4$ with gauge group $G$ and coupling
constant $\tau$ is equivalent to  SYM ${\cal N}=4$ with dual gauge group
$^LG$ \cite{GNO} and dual coupling constant $^L\tau$ , with
\beq
^L\tau=-\frac{1}{\tau}
\eeq
for simply laced algebras, where
\beq
\tau=\frac{\theta}{2\pi}+\frac{4\pi i}{g_{4d}^2},\qquad ^L\tau=
\frac{^L\theta}{2\pi}+\frac{4\pi i}{(^Lg_{4d})^2}.
\eeq
The symmetry has to be understood as an operator isomorphism between
the two theories \cite{kapu}. Since it interchanges electric and 
magnetic charges, it maps a Wilson operator \cite{wil} onto a 't Hooft 
operator \cite {thoo} and vice-versa.
Conjectures have also been suggested for chiral primary 
operators \cite{intri}, surface operators \cite{gw} and domain walls 
\cite{gawi}.

An advance has recently been made with \cite{got} where the 
conjecture
has been extended to correlation functions of gauge invariant operators.
The set of observables in SYM ${\cal N}=4$ are related by the S-duality
requirement
\beq
\langle \Pi_i {\cal O}_i\rangle_{G,\tau}=\langle \Pi_i {^L\cal O}_i
\rangle_{^LG,^L\tau}.
\eeq
This property is both interesting and difficult to prove since it
involves strong coupling calculations. The choice has been focused
on a 't Hooft operator $T(^LR)$ in a theory with gauge group $G$,
$^LR$ being a representation of the dual group $^LG$.

The expectation value of a 't Hooft loop can be computed by a 
path-integral where the integration is performed over all fields 
which have a prescribed singularity along the loop. In the weak coupling
regime quantum fluctuations around the classical monopole configuration 
can also be obtained up to one loop order and
a recipe has been provided to compute the loop perturbatively
at any desired higher order.

This result has subsequently been compared with a strong coupling 
calculation of a Wilson loop with dual
gauge group and dual coupling (see (\ref{dual})).
\vskip.5truecm

To compute Wilson loops where some fractions of supersymmetries
are preserved, one may resort to matrix models where
explicit calculations are feasible, as conjectured in \cite{esz,dg}
and proved in \cite{pes}. 
A rather interesting family of contours can be obtained by coupling
three of the six scalars and by restricting the contours to lie on
a great $S^2$ inside $S^3$. The related 1/8 BPS loop operators
are conjectured to correspond to the ``zero-instanton sector''
of the two-dimensional Yang-Mills theory ($YM_2$) on $S^2$ 
\cite{dgrt}. In turn this was proved long ago to be equivalent to 
a Gaussian matrix model with area dependent coupling $g^2A=-2g_{4d}^2$ 
\cite{SK},\cite{bagr}.
Several results which comply with this conjecture have appeared 
recently in \cite{bal}.
 
From matrix models a strong coupling expression for the Wilson loop
can be extracted, to be compared with the weak coupling expression
of the 't Hooft loop hitherto obtained. This can eventually be
used to test the
S-duality conjecture
 
\beq
\label{dual}
\langle T(^LR)\rangle_{G,\tau}=\langle W(^LR)\rangle_{^LG,^L\tau}.
\eeq

An even bolder conjecture has been proposed in \cite{gp}. 
After retrieving the correspondence between a (supersymmetric)
Wilson loop in SYM $\cal N$=4 and the zero-instanton sector of
the loop in $YM_2$, the authors extended this relation to 
suitable 't Hooft
operators. More precisely they suggested that the expectation value
of the 1/2 BPS circular 't Hooft loop in representation $^LR=
(m_1,...,m_N)$ in SYM $\cal N$=4 with gauge group $G$ and with an
imaginary coupling ($\theta=0$) could be obtained
from the partition function ${\cal Z}$ of $YM_2$ with gauge group $G$ 
around an unstable instanton \cite{Wi} labelled by $^LR$

\beq
\label{YM2}
\langle T_{^LR}({\cal C})\rangle_{G,\tau}=\frac{{\cal Z}(g;m_1,...,m_N)}
{{\cal Z}(g;0,...,0)},
\eeq
where the configuration $(m_1,\cdots,m_N)$ is related to the boxes
in the Young tableau.

Similarly, correlation functions of the 1/2 BPS 't Hooft loop with
any number of 1/8 BPS Wilson loops inserted on the $S^2$ linked to
the 't Hooft loop, could be computed in $YM_2$ by calculating the
Wilson loop correlation functions around a fixed unstable instanton.
 
These suggestions are particularly intriguing since they point towards
endowing those instantonic sectors with a ``physical'' meaning.

In fact in \cite{gp} the check was limited to 
the K-antisymmetric
representations of the gauge group $U(N)$, which cannot be screened to
give rise to sub-leading saddle points in the path integral localization
(the ``monopole bubbling'' phenomenon \cite{KW}). Moreover the
choice of $U(N)$ hid the possible occurrence of different
representations $R$ and $^LR$ in the general case, $U(N)$ being self-dual.

Our purpose in this paper is to extend the analysis to the gauge
group $SU(N)$ and to its dual $SU(N)/Z_N$.

In Sect.2 we develop the harmonic analysis in $SU(N)$ and $SU(N)/Z_N$
of the partition function and of a Wilson loop.
We remark that the Poisson transformation, which is the bridge between 
the expansions in terms of characters and of unstable instantons 
respectively, provides us with two {\it different} expressions for the
{\it same} quantity.

In Sect.3 we test the conjecture of ref.\cite{gp} of a relation
between a Wilson loop in the $K$-fundamental representation of SU(N) and a
't Hooft loop, obtained by singling out in the partition function
the contribution of an instanton belonging to the same representation.
The test was successfully performed in \cite{gp} for the group $U(N)$.
The novelty in our case is that the $K$-irrep is not present in
$SU(N)/Z_N$. As a consequence the test can only be exploited starting
from $SU(N)/Z_N$ for the partition function and ending 
in $SU(N)$.
Then we discuss the case of the adjoint representation. In this case both
$SU(N)$ and $SU(N)/Z_N$ are viable. However it turns out that the
instanton contribution to the partition function, which should 
correspond to a 1/2 BPS 't Hooft loop in SYM ${\cal N}=4$, 
indeed presents  some 
extra terms (subleading in $N$) with respect to the Wilson loop in
the same representation. This is a concrete realization of the possibility
mentioned in \cite{gp} and there interpreted as a subleading contribution 
in the path-integral localization of SYM ${\cal N}=4$.

Limitations occurring when considering correlators between Wilson loops
and a 't Hooft loop are also briefly pointed out.

Finally Sect.4 contains our conclusions together with some insight into 
possible future developments.

\section{The harmonic analysis on $SU(N)$ and $SU(N)/Z_N$}

The basic ingredient in computing the partition function
and Wilson loop correlators in $YM_2$ with gauge group SU(N) 
is the heat kernel 
on a two-dimensional
cylinder ${\cal K}(A;U_2,U_1)$ of area $A=L\tau$ ($L$=base circle,
$\tau$ = length), and fixed holonomies at the
boundaries $U_1$ and $U_2$. The only geometrical dependence of the
kernel is on its area, thanks to the invariance of $YM_2$ under 
area-preserving diffeomorphisms \cite{Wi}. The kernel enjoys the basic
sewing property
\beq
\label{heat}
{\cal K}(L\tau:U_2,U_1)=\int dU(u) {\cal K}(Lu;U_2,U(u)){\cal K}
(L(\tau-u));U(u),U_1).
\eeq 

The partition function on a sphere with area $A$ is expressed
as ${\cal K}(A;{\bf 1},{\bf 1})$.

The kernel ${\cal K}$ can be expanded as a series of the characters 
$\chi_{R}$
of all the irreducible representations (irreps) of SU(N), 
according to the equation
\beq
\label{expa}
{\cal K}(A;U_2,U_1)=\sum_R \chi^{\dagger}_R(U_2)\chi_R(U_1)\exp\Big[
-\frac {g^2A}{4}C^R\Big],
\eeq
$C^R \equiv C_2(R)$ being the quadratic Casimir operator of the R-
representation.

\vskip.3truecm

Now we move our interest to the group $SU(N)/Z_N$, $Z_N$ being the
center of $SU(N)$. The elements of $Z_N$ are the roots of unity
$z=\exp\frac{2\pi i n}{N},\,\,\, n=0,\cdots,N-1.$ The homotopy
of $SU(N)/Z_N$ is non trivial; the bundles of $SU(N)/Z_N$ over
a two-dimensional closed oriented Riemann surface $\Sigma$ can
be topologically classified by the choice of a value of $z$ \cite{gpss}, 
\cite{bgv}.

To compute the partition function of $YM_2$ (and, more generally,
its basic propagation kernel) one has to sum over the topologies
of those bundles. A convenient way to do so is to weight the contribution
of each sector with a representation $\chi_k(z)$ of $Z_N$, obtaining the
following refined expression
\beeq
\label{proj}
\K_k(A;U_2,U_1)&=& \sum_{z\in Z_N} z^k \K(A;zU_2,U_1) \\ \no 
&=&\sum_{n=0}^{N-1} \sum_R  e^{\frac{2\pi i n}{N}(k-m^R)}
\chi^{\dagger}_R(U_2)\chi_R(U_1)\exp\Big[
-\frac {g^2A}{4}C^R\Big],
\eeeq
where
the holonomy $U_2$ has been ``twisted'' and $m^{(R)}=\sum_{q=1}
^{N-1} m^{(R)}_q$ is the total number of boxes of the Young 
tableau. 

In this equation $k$ selects a sector of irreps of $SU(N)$ by the
rule $k(R)= m^{(R)}$. When $k=0$ the irreps of $SU(N)$ are ``neutral''
and thereby belong to $SU(N)/Z_N$.

Choosing $U_1=U_2={\bf 1}$, the contribution of the k-sector
to the partition function takes the expression
\beq
\label{part}
\Z_k(A)=\sum_R (d_R)^2 \exp\Big[-\frac{g^2A}{4}C^R\Big]\delta
_{[N]}(k-m^{(R)}),
\eeq
where $d_R$ is the dimension of the R-irrep and $\delta_{[N]}$ is 
the $N$-periodic delta-function.

Summing over $k$, the $SU(N)$ partition function is immediately
recovered. The sector $k=0$ provides instead the partition function
$\Z_0(A)$ of $SU(N)/Z_N$, where the contributions from different
topological bundles are summed over (see eq.(\ref{inve})).

Introducing the explicit expression for the characters \cite{HW} 
enables us
to write eq.(\ref{part}) explicitly in terms of a new set of
indices $\{l_i\}=(l_1, \ldots, l_N),\,l_i=m_i+N-i$ (see
the Appendix). By recalling the relations
\beeq
\label{casimiri}
C_2(R)&=&
\sum_{i=1}^{N} \Big( l_{i}-\frac{l}{N} \Big)^2-
\frac{N}{12}(N^2-1)
\nonumber \\
d_{R}&=&\Delta(l_1,...,l_{N}),\qquad l=\sum_{i=1}^{N}l_{i},
\eeeq
where $\Delta$ is the Vandermonde determinant , we get \cite{bgv}
\beeq
\label{partip}
&&\Z_k (A)=\frac{(2\pi)^{N-1}}{N!\,\sqrt\pi } \sum_{l_i=-\infty}^
{+\infty} \int_0^{2 \pi}
d\alpha\, e^{-\Big(\alpha - \frac{2\pi}N l \Big)^2}
\delta_{[N]} \Big( k- l + \frac{N(N-1)}2 \Big) \nonumber \\
%&& \ph{\Z_k (A)=\frac1{N!\,\sqrt\pi } \sum_{l_i=-\infty}^{+\infty} } 
&\times& \quad
\exp \left [ -\frac{g^2 A}{4} C_2 (l_i)\right ]
\Delta^2(l_1,...,l_N).
\eeeq

The dual representation in this context is realized by means of a 
Poisson transformation
\begin{eqnarray}
\label{poisson}
&&\sum_{l_i=-\infty}^{+\infty}F(l_1,\ldots,l_N)=
\sum_{n_i=-\infty}^{+\infty}\tilde F(n_1,\ldots,n_N),\no \\
&&\tilde F(n_1,\ldots,n_N)=\int_{-\infty}^{+\infty}dz_1\ldots dz_N
F(z_1,\ldots,z_N)\exp\Bigl[2\pi i(z_1n_1+\ldots+z_Nn_N)\Bigr].
\end{eqnarray}
In order to perform this multiple Fourier transform,
we remember that the transformation of a product is turned into
a convolution; moreover we recall the result 
\begin{eqnarray}
\label{ftrans}
&&\int_{-\infty}^{+\infty}dz_1\ldots dz_N
\exp\Bigl[i(z_1p_1+\ldots+z_Np_N)\Bigr]\Delta(\{z_i\})
\exp \Big( -\frac{g^2A}{8}\sum_{q=1}^{N}z_q^2\Big)= \no \\
&&\Big[\frac{4i}{g^2A}\Big]^{\frac{N(N-1)}{2}}\Big[\frac{8\pi}{g^2A}
\Big]^{\frac{N}{2}}\Delta(\{p_i\})
\exp\Big(-\frac{2}{g^2A}\sum_{q=1}^{N}p_q^2\Big).
\end{eqnarray}
Taking these relations into account, eq.~(\ref{partip}) becomes
\begin{equation}
\label{sectors}
{\cal Z}_k(A)=\sum_{n=0}^{N-1}\exp\Big[\frac{2\pi ink}{N}\Big]
{\cal Z}^{(n)}(A),
\end{equation}
where
\begin{equation}
\label{nsector}
{\cal Z}^{(n)}(A)=(-1)^{n(N-1)}\,{\cal C}(A,N)\sum_{n_q=-\infty}^{+\infty}
\delta(n-\sum_{q=1}^{N}n_q)
\exp\Big[-\frac{4\pi^2}{g^2A}\sum_{q=1}^{N}(n_q-\frac{n}{N})^2\Big]
\zeta_n(\{n_q\}),
\end{equation}
%with $${\cal C}(A,N)=\frac{(2\pi)^{2N-3}}{N!}\sqrt{\pi N} \,e^{\frac{g^2A%}{48}
%N(N^2-1)} \,2^{N(N+\frac{1}{2})}\, (g^2A)^{-(N^2-N/2-1/2)}$$ and
with
\beeq
\label{zeta}
&&\zeta_n(\{n_q\})=\int_{-\infty}^{+\infty}dz_1\ldots dz_N \exp\Big[-
\frac{1}{2}\sum_{q=1}^{N} z_q^2\Big] \Delta ( \{
\sqrt{\frac{g^2A}{2}} z_q+2 \pi n_q\} )
\Delta(\{\sqrt{\frac{g^2A}{2}}z_q-2\pi 
n_q\})\no \\
&=& \int_{-\infty}^{+\infty}dz_1\ldots dz_N \exp\Big[-
\frac{1}{2}\sum_{q=1}^{N} z_q^2\Big] \Delta^2 ( \{
\sqrt{\frac{g^2A}{2}} z_q-2 \pi i n_q\} )
\eeeq
and ${\cal C}(A,N)$ an unessential normalization factor \cite
{bgv}. $\Z^{(n)}$ is clearly invariant under a common translation
$\{n_q\}\to \{n_q-h\},\,\,h\in {\bf Z}$ : $\Z^{(n)}=\Z^{(n+hN)}$.

The classical instanton action ${\cal S}=\Big[\frac{4\pi^2}{g^2A}
\sum_{q=1}
^{N}(n_q-\frac{n}{N})^2\Big]$ can be nicely compared to the Casimir
expression in the exponential of eq.(\ref{part}). One can already
remark that the factor $\frac{4\pi^2}{g^2A}$ here corresponds to
the factor $\frac{g^2 A}{4}$ there, as suggested by duality.

The duality can most easily be
realized by taking the sum over the sectors $k$, firstly in (\ref{part}):
\beq
\label{av}
\sum_{k=0}^{N-1} \Z_k= \sum_R (d_R)^2 \exp\Big[-\frac{g^2A}{4}
C_2(R)\Big],
\eeq
as expected in $SU(N)$ (the $\delta$-constraint on $m^{(R)}$ has 
disappeared) and then in (\ref{sectors}):
\beq
\label{av1}
\Z=\sum_{k=0}^{N-1} \Z_k=\Z^{(0)}={\cal C}(A,N)
\sum_{n_q=-\infty}^{+\infty}
\delta_{[N]}(\sum_{q=1}^{N}n_q)
\exp\Big[-\frac{4\pi^2}{g^2A}\sum_{q=1}^{N}n_q^2\Big]
\zeta_n(\{n_q\}).
\eeq

The dual relation can easily be obtained from eq.(\ref{sectors})
\beq
\label{inve}
\sum_{n=0}^{N-1}\Z^{(n)}=\Z_0.
\eeq

The expressions (\ref{av1}) and (\ref{inve})
are indeed symmetric under the interchange
of the two sets of integers $\{m_q\}$ and $\{n_q\}$. 

\vskip.5truecm

The next step to be performed is to obtain the quantum average of a
Wilson loop in $SU(N)/Z_N$. In so doing we should confine ourselves
to the set of ``neutral'' representations for the loop ($\sum m_i=0,
\,\,mod\, N$), otherwise the quantum expression would involve different 
sectors of $SU(N)$. 

Let us therefore consider a regular non self-intersecting loop placed 
on the equator of our sphere $S^2$
\beq
\label{wloop}
\W_0(\frac{A}{2},\frac{A}{2})= \frac1{\Z_0}\sum_{z\in Z_N}\int dU
\K(\frac{A}{2};z\cdot {\bf 1}, U)\,\frac{1}{d_0}Tr_0[U]\,
\K(\frac{A}{2};U,{\bf 1}).
\eeq
\noindent
If we choose the adjoint representation, introducing characters, we get

\beeq
\label{waloop}
\W_{adj}(\frac{A}{2},\frac{A}{2})&=&\frac1{\Z_0 \, (N^2-1)} 
\sum_{R,S} d_{R}d_{S}
\exp\left[-\frac{g^2 A}{8}(C^{(R)}+C^{(S)})\right]
\no \\
&\times & \int dU \, {\rm Tr}_{adj}[U]
\, \chi_{R}(U) \chi_{S}^{\dagger}(U)\,
\delta_{[N]}( m^{(S)})\, .
\eeeq

In the $0$-sector the loop exhibits the expected $\delta_{[N]}$
constraint on the total number of boxes $m^{(S)}$ of the Young tableau.

By making the expression of the characters explicit, after integrating 
over the group variables, taking suitable invariance under 
permutations into account and invariance of the Vandermonde
determinants under constant translations in their arguments, 
a calculation (partially sketched in the Appendix; see also \cite{bgv}) 
leads to
\beeq
\label{wilsonint}
\W_{adj} (\frac{A}{2},\frac{A}{2})&=&\frac1{N+1}
\Biggl\{ 1+ 
\frac2{\Z_0 \, (N-1)}\,
\frac{(2\pi)^{N-1}}{\sqrt \pi \,N!}
\sum_{l_i=-\infty}^{+\infty} \,
\sum_{1=q_1<q_2}^{N} \exp [ - \frac{g^2 A}4 ( l_{q_2}
-l_{q_1}+1)]  
\no \\
&\times& 
\int_0^{2 \pi}
d\alpha \,\,
e^{-( \alpha - \frac{2\pi}N l  )^2} 
\delta_{[N]} ( - l + \frac{N(N-1)}2 )  \no \\
&\times& 
\Delta(l_1,...,l_N) \, \Delta(l_1,\ldots ,  l_{q_1} -1,
\ldots,   l_{q_2}+1 , \ldots, l_N)  
\Biggr\}  
\,,
\eeeq

\vskip.5truecm
Before undertaking the Poisson transformation it is useful to factorize
the $\delta_{[N]}$-constraint using its exponential representation
$\delta_{[N]}(q)=\frac1N \sum_{p=0}^{N-1}e^{\frac{2\pi i\,p\,q}{N}}.$
Then, by repeating the procedure used for  ${\cal Z}_0$, a long 
but straightforward calculation leads to
\begin{equation}
\label{wsector}
{\cal W}_{adj}(\frac{A}2,\frac{A}2)=\frac{1}{N+1}+\frac{1}{{\cal Z}_0}
\sum_{n=0}^{N-1}
{\cal W}^{(n)}(\frac{A}2,\frac{A}2)
\end{equation}
where
\beeq
\label{nwilson}
{\cal W}^{(n)}&=&(-1)^{n(N-1)}\, \frac{2\,{\cal C}(A,N)}{N^2-1}
\sum_{r<s}
\sum_{n_q=-\infty}^{+\infty}\delta(n-\sum_{q=1}^{N}n_q)
\no \\
&\times& 
\exp [-\frac{4\pi^2}{g^2A}\sum_{q=1}^{N}(n_q-\frac{n}{N})^2 ]
\exp [i\pi (n_s-n_r)]\Omega_n(\{n_q\})
\eeeq
and
\beeq
\label{fluct}
&&\Omega_n(\{n_q\})=\int_{-\infty}^{+\infty}dz_1\ldots dz_N
\exp[ -\frac{1}{2}\sum_{q=1}^{N}z_q^2] \exp[\frac{i}{2}
\sqrt{\frac{g^2A}{2}}(z_r-z_s)]
\times \no \\
&&\Delta(\sqrt{\frac{g^2A}2}z_1-2\pi n_1,.,
\sqrt{\frac{g^2A}2}z_N-2\pi n_N)\,
\Delta(\sqrt{\frac{g^2A}2}z_1+2\pi n_1,.,
\sqrt{\frac{g^2A}2}z_N+2\pi n_N).
\eeeq

\vskip.5 truecm

Obviously the limitation of considering only ``neutral'' representations 
for the Wilson loop does not concern the group $SU(N)$.
If we choose for example the $K$-antisymmetrical fundamental 
representation $(1,\cdots,1,0,\cdots,0)$ we get
\beeq
\label{wKloop}
\W(\frac{A}{2},\frac{A}{2})&=&\frac1{\Z \, d_K} 
\sum_{R,S} d_{R}d_{S}
\exp\left[-\frac{g^2 A_1}{8}(C^{(R)}+C^{(S)})\right]
\no \\
&\times & \int dU \, {\rm Tr}_K[U]
\, \chi_{R}(U) \chi_{S}^{\dagger}(U),
\eeeq
and, by repeating the technical procedures we followed in the 
case of the adjoint representation, we obtain the expression
\beeq
\label{long}
\W(\frac{A}{2},\frac{A}{2}) &=& \frac1{\Z} \sum_{l_i=-\infty}^{+\infty}\int_{-\infty}
^{+\infty} d\beta
\int_{-\infty}^{+\infty}dl
\, e^{i\beta(l-\sum_il_i)}  \no \\
&&\int_0^{2\pi}\, d\alpha\, e^{-(\alpha-\frac{2\pi l}{N})^2}\exp
\Big[{-
\frac{g^2 A}{8} \Big(2C(l_i)-2\sum_{j=1}^K l_j+\frac{K}{N}(N+2l-K)\Big)
}\Big]
\no \\
&&\Delta(l_1-1,l_2-1,\cdots,l_K-1,l_{K+1},\cdots,l_N) 
\Delta(l_1,l_2,\cdots,l_N).
\eeeq 
After a Poisson resummation, we finally reach its expansion in terms of
instantons
\beeq
\label{WKs}
&&{\cal W}(\frac{A}{2},\frac{A}{2})=\frac1{\Z} e^{\frac{g^2AK^2}{16N}}
\sum_{\{n_i\}}\delta_{[N]}(\sum_{i=1}^N n_i) 
e^{i\pi\sum_{j=1}^{K}n_j}
\int_{-\infty}^{+\infty}dy_1\cdots dy_N \Pi_{i<j}\Big[4\pi^2 n_{ij}^2-
y_{ij}^2\Big]\no \\
&& \exp{\Big[-\frac{4\pi^2}{g^2A}\sum_j n_j^2\Big]}
e^{-\frac{i}2\sum_{j=1}^{K}y_j}e^{-\frac1{g^2A}\sum_j y_j^2}.
\eeeq

\section{The conjecture}

As discussed in the Introduction, the average value of a 1/2 BPS
 t'Hooft circular
loop winding on a large circle on $S^2$ in SYM ${\cal N}$=4
with gauge group $G$  in the representation $^LR= (m_1,\cdots,m_N)$
has been conjectured to be obtained from the contribution to 
the partition function 
$\Z$ of $YM_2$ of an
unstable instanton labelled by $^LR$ (see eq.(\ref{YM2})). In turn this 
should be dual to the ``zero instanton'' contribution to the average 
value of a Wilson loop in the $^LR$ representation (in the
character expansion) of the group $^LG$, winding over a large
circle of $S^2$ of $YM_2$ $\,$(\cite{gp}).

In the following, the possibility of singling out different $k$-sectors of
$SU(N)$ will not be pursued. In fact we think nobody knows at present the 
relevance (if any) of those sectors in the correspondence $YM_2
\leftrightarrow SYM\, {\cal N}=4$, in particular the meaning of the
counterpart (if any) of the $k$-parameter of $YM_2$ in the
$SYM\, {\cal N}=4$ context.

We are now in the position to discuss the conjecture when the groups
considered are $SU(N)$ and its dual $SU(N)/Z_N$. Let us first start
from the character expression of $\Z_0$ moving to its
dual instanton expansion in $SU(N)$.

The ``zero instanton'' contribution
of $\Z_0$ is easily 
derived from 
eq.(\ref{sectors})
\beq
\label{Zzero}
\Z_0^{[0]}= \int dz_1,\cdots,dz_N \exp \Big[-\frac12 \sum_q z_q^2\Big] \Delta^2(\{z_q\}),
\eeq
where the normalization has been suitably modified.

As a first example, we  calculate the instanton
contribution to the partition function $\Z_0$ corresponding to the
$K$-fundamental representation $\{n_q\}=(1,\cdots,1,0,\cdots,0)$
with the first $K$-elements being unity, and permutations thereof.
  
Inserting this configuration in eqs.(\ref{sectors}),(\ref{nsector})
and (\ref{zeta}), we get
\beq
\label{zetaK}
\Z_0^{[K]}=(-1)^{K(N-1)}
e^{\frac{4\pi^2K^2}{g^2AN}}
\int_{-\infty}^{+\infty}dz_1\cdots dz_N \Pi_{i<j}\Big[z_{ij}^2\Big]
e^{-2\pi i \sqrt{\frac2{g^2A}}\sum_{j=1}^{K}z_j}\,\,e^{-\frac12 
\sum_j z_j^2},
\eeq
where permutations have been taken into account.

According to the conjecture, this result is to be compared to the 
zero-instanton contribution 
in eq.(\ref{WKs}). 
The change of variables $y_i=\sqrt{\frac{g^2A}2}z_i$ 
would lead to a perfect agreement with eq.(\ref{zetaK}) under the
interchange $\frac{8\pi^2}{g^2A}\leftrightarrow \frac{g^2A}8$, 
were it not for 
the sign factor in (\ref{zetaK}). The occurrence of a similar factor 
was also noticed in \cite{gp}.

\vskip.5truecm

The other option ($SU(N)\to SU(N)/Z_N$) is not viable. As a matter of 
fact the presence of the constraint $\delta_{[N]}(\sum_{q=1}^{N}n_q)$ 
in $\Z^{(0)}$ makes the representation  $\{n_q\}=(1,\cdots,1,0,
\cdots,0)$ 
for the 't Hooft loop impossible, as it is not shared by the group
$SU(N)/Z_N$ (see eq.(\ref{av1})).

\vskip.3truecm

At this point some comments concerning other irreps are in order.

Suppose we consider a 't Hooft loop in the adjoint representation.
The total number of boxes in the Young tableau being $N$ in this case, 
we can equally well consider $SU(N)$ or $SU(N)/Z_N$.
Going back to eqs.(\ref{YM2}) and (\ref{sectors}), the Young tableau
of the adjoint representation has the configuration
 $\{n_q\}= (2,1,\cdots,1,0)$, which is equivalent $ mod \, N$ to
$(1,0,\cdots,0,-1)$, its highest weight. We get
\beq
\label{adjinst}
\Z_{adj}=\int_{-\infty}^{+\infty} dz_1,\cdots,
dz_N \exp \Big[-\frac12 \sum_{q=1}^{N} z_q^2\Big] 
 \exp \Big[\frac{2\sqrt{2}\pi i}{\sqrt{g^2A}}z_{1N}\Big]\Delta^2(\{z_q\}).
\eeq
Taking invariance under permutations into account, it becomes
\beeq
\label{perm}
&&\Z_{adj}= \Big(1+\frac{1}{N}\Big) \int_{-\infty}^
{+\infty} dz_1,\cdots,dz_N 
\Delta^2(\{z_q\})
\exp \Big[-\frac12 \sum_{q=1}^{N} z_q^2\Big] \times \no \\
&&\Big[\frac{\sum_{r,s=1}^{N}\exp(2\pi i\sqrt{2/g^2A}z_{rs})-1}{N^2-1}
-\frac1{N+1}
\Big]\no \\
&=& \Big[ (1+\frac1{N}) \int {\cal D}F \exp(-\frac12 
TrF^2)
\frac{1}{N^2-1}\Big(|Tr[\exp\big(2\pi i \sqrt{\frac{2}{g^2A}}F\big)]|^2
-1\Big)\Big]-
\frac{\Z^{[0]}}{N}.
\eeeq 
Here $F$ is a traceless hermitian matrix.
\vskip.5truecm
The ``zero instanton'' contribution
to the Wilson loop in the adjoint representation can easily be obtained
from eq.(\ref{wsector})
\beeq
\label{wzero}
&&\W_{adj}^{[0]}(\frac{A}2,\frac{A}2)=\frac{1}{N+1}\Bigg[1+
\frac{N}{\Z^{[0]}}\int_{-\infty}^{+\infty}dz_1\ldots dz_N
\exp\Big[ -\frac{1}{2} \sum_{q=1}^{N}z_q^2\Big] \times \no \\ 
&&\exp\Big[\frac{i}{2}
\sqrt{\frac{g^2A}{2}}z_{12}\Big]
\Delta^2(z_1,
\ldots,z_N)\Bigg].
\eeeq
Eventually 
the expression above turns into the matrix integral \cite{tony}
\beq
\label{mint}
\W^{[0]}_{adj}=\frac{1}{\Z^{[0]}}\int {\cal D}F \exp(-\frac{1}{2} TrF^2)
\frac{1}{N^2-1}\big(|Tr[\exp\frac{ig}{2}\sqrt{\frac{A}{2}}F]|^2-1\big).
\eeq

\vskip.5truecm

Comparing eqs.(\ref{perm}) and (\ref{mint}), 
we notice the expected duality relation $\frac{g^2A}{8}
\leftrightarrow \frac{8\pi^2}{g^2A}$, but also the occurrence in
(\ref{perm}) of extra
terms, possibly related to the afore mentioned "monopole bubbling" 
\cite{KW}.

We end this Section with a comment concerning correlators. In a theory
with gauge group $G$, Wilson loops are labelled by irreps of $G$,
whereas 't Hooft loops are labelled by irreps of $^LG$. As a consequence,
in the case $SU(N)\leftrightarrow SU(N)/Z_N$, in a correlator
$<W(R)T(^LR)>$ one cannot choose totally antisymmetric representations 
for
both $R$ and $^LR$ since one of the two representations is unavailable
(see eq.(\ref{WKs})).
Antisymmetric-adjoint and adjoint-adjoint would be viable choices,
but possible subleading contributions would be involved.

\section{Conclusions}

We have extended the conjecture of ref.\cite{gp} concerning a 1/2 BPS
't Hooft loop in the group $U(N)$, to the more general case of a group
which is not self-dual. We have concretely examined the choice $SU(N)
\leftrightarrow SU(N)/Z_N$. The duality mapping is performed
in our treatment by a Poisson transformation between an expansion in
terms of characters and the one in terms of instantons. 

The novelty in the case of groups which are not self-dual lies
in the circumstance that not all representations are shared by them.
For instance it is well known 
that the  spinorial representations of $SU(2)$  are not shared by its 
dual partner $SU(2)/Z_2$.  

In the example $SU(N) \leftrightarrow SU(N)/Z_N$
we have discussed, if we want a 't Hooft loop belonging to one
of the fundamental irreps of $SU(N)$, we ought to start from
$SU(N)/Z_N$, landing, after the Poisson transformation, in $SU(N)$.

We have also briefly discussed the adjoint irrep, which belongs to
both $SU(N)$ and $SU(N)/Z_N$. Here we have concretely realized
that this choice in the partition function for the 't Hooft loop
involves subleading corrections, as expected on general grounds 
\cite{gp}.

When considering correlators between Wilson loops and a 't Hooft loop
according to the conjecture, possible subleading saddle point 
contributions
are involved, because ``minimal'' representations for both are impossible.

It would be nice in the future to be able to extend the conjecture
beyond the 1/2 BPS 't Hooft loop. As a preliminary requirement
we need to thoroughly understand  more
general configurations of a 't Hooft loop in SYM ${\cal N}$=4,
in particular their contributions as saddle points in the
localization of the path-integral \cite{pes}.

On more general grounds one might speculate whether topologically
inequivalent $k$-sectors of $SU(N)$ in $YM_2$ would possess via
the conjecture any counterpart in the form of some peculiar properties 
of $SYM \, {\cal N}=4$.

From the mathematical side one should perhaps understand in
a more general and systematic way the connection between
a formulation of duality in terms of algebras and of groups.
We remark
that previous treatments were mostly based on a relation between
algebras exchanging their highest weights under the duality 
transformation \cite{GNO},\cite{kapu},\cite{got}. Here the conjecture 
forces us to choose their group counterparts where duality operates
in the form of an integral Poisson transformation. 

\vskip.3truecm

{\bf ACKNOWLEDGEMENTS}

We thank Luca Griguolo for useful discussions and Simone Giombi
for a fruitful correspondence.

\vskip.3truecm

\section{Appendix}
Let us introduce for $SU(N)$ the usual variables 
\beq
\label{seq}
\hat l_q=m_q+N-q, \qquad\qquad q=1,\cdots,N-1,
\eeq
which give rise to a strongly monotonous sequence $\hat l_1>
\hat l_2>\cdots,\hat l_{N-1}>0$ \cite{bgv}. Then, with the twofold 
purpose of
extending the range of the $\hat l_q$'s to negative integers 
and of gaining the symmetry over permutations of a full set of $N$ 
indices, we introduce the obvious equality
\beq
\label{obvious}
\sqrt{\pi}=\int_0^{2\pi}d\alpha \sum_{\hat l_N=-\infty}^{+\infty}
e^{-(\alpha-\frac{2\pi}{N}\sum_{j=1}^{N-1}\hat l_j-2\pi\hat l_N)^2},
\eeq
where $\hat l_N$ is a dummy quantity. Now we extend the representation 
indices by defining the new set
\beeq
\label{newi}
&&l_q=\hat l_q + \hat l_N,\qquad\qquad q=1,\cdots,N-1, \no \\
&&l_N=\hat l_N,
\eeeq 
which appears in eq.(\ref{casimiri}) and  the equations that follow.
\vskip.5truecm
In terms of these indices eq.(\ref{waloop}) takes the form
\beeq
\label{wilsonfin}
&&\frac1{N^2-1}+ \W_{adj}(\frac{A}{2},\frac{A}{2})
=\frac1{\Z_0 \, (N^2-1)}
\sum_{l_i^R,\, l_i^S=-\infty}^{+\infty} 
\frac1{2\pi^2 (N!)^2}
\sum_{n=-\infty}^{+\infty}
\int_0^{2\pi} d\theta_1 \ldots d \theta_N \, \no \\
&&\times
\int_0^{2 \pi}
d\alpha_1 \,d\alpha_2 \, \, e^{ -\left( \alpha_1 - \frac{2\pi}N l^R 
+ 2\pi n\right)^2}\,
e^{-\left(\alpha_2 - \frac{2\pi}N l^S \right)^2 }
\exp\left[-\frac{g^2 A}{8}\left(C_2(l^R_i)+C_2(l^S_i)\right)\right] \\
&&\times \sum_{p,q=1}^{N} e^{i (\theta_p-\theta_q)}\;
\prod_{h=1}^N e^{i l_h^R \theta_h} \prod_{r=1}^N e^{-i l_r^S \theta_r} \,
\delta_{[N]} \left( - l^S + \frac{N(N-1)}2 \right) 
\Delta(l^R_1,...,l^R_N) \Delta(l^S_1,...,l^S_N) \no\,.
\eeeq
By making the expression of the characters explicit, by taking invariance 
under permutations into account, and after integrating over the group 
variables, eq.(\ref{wilsonint}) is eventually recovered.

\end{document}